\documentclass[12pt]{article}
\usepackage{amsmath}
\usepackage{bm}
\usepackage{color}

\begin{document}
\begin{center}
\begin{large}
{\bf System of interacting harmonic oscillators in rotationally invariant noncommutative phase space}
\end{large}
\end{center}

\centerline {Kh. P. Gnatenko \footnote{E-Mail address: khrystyna.gnatenko@gmail.com}}
\medskip
\centerline {\small \it Ivan Franko National University of Lviv, Department for Theoretical Physics,}
\centerline {\small \it 12 Drahomanov St., Lviv, 79005, Ukraine}
\centerline {\small \it  Laboratory for Statistical Physics of Complex Systems}
\centerline {\small \it  Institute for Condensed Matter Physics, NAS of Ukraine Lviv, 79011, Ukraine}

\abstract{Rotationally invariant space with noncommutativity of coordinates and noncommutativity of momenta of canonical type is considered. A system of $N$ interacting harmonic oscillators in uniform filed and a system of $N$ particles with harmonic oscillator interaction are studied. We analyze effect of noncommutativity on the energy levels of these systems. It is found that influence of coordinates noncommutativity on the energy levels of the systems increases with increasing of the number of particles.  The spectrum of $N$ free particles in uniform field  in rotationally-invariant noncommutative phase space is also analyzed. It is shown that the spectrum corresponds to the spectrum of a system of $N$ harmonic oscillators with frequency determined by the parameter of momentum noncommutativity.

}

\section{Introduction}

Recently much attention has been devoted to studies of a quantum space realized on the basis of idea that the spatial coordinates might be noncommutative.  The noncommutative space of canonical type has been  studied intensively. In the space the coordinates satisfy the following commutation relations
  \begin{eqnarray}
[X_{i},X_{j}]=i\hbar\theta_{ij},\label{form101}{}
\end{eqnarray}
where $\theta_{ij}$ are elements of constant antisymmetric matrix, parameters of coordinate noncommutativity.  In noncommutative phase space the momenta are supposed to be noncommutative too. The commutation relations read
  \begin{eqnarray}
[P_{i},P_{j}]=i\hbar\eta_{ij}.\label{form1001}{}
  \end{eqnarray}
Commutation relations for coordinates and momenta are generalized as
    \begin{eqnarray}
[X_{i},P_{j}]=i\hbar(\delta_{ij}+\gamma_{ij}).\label{form10001}{}
  \end{eqnarray}
where $\eta_{ij}$, $\gamma_{ij}$ are elements of constant matrixes.

 Coordinates $X_i$ and momenta $P_i$ which satisfy  (\ref{form101}), (\ref{form1001}) can be represented as
\begin{eqnarray}
X_{i}=x_{i}-\frac{1}{2}\sum_{j}\theta_{ij}{p}_{j},\label{repx}\\
P_{i}=p_{i}+\frac{1}{2}\sum_{j}\eta_{ij}{x}_{j},\label{repp}
\end{eqnarray}
here  $x_i$, $p_i$ are coordinates and momenta which satisfy
\begin{eqnarray}
[x_{i},x_{j}]=0,\label{oc}\\{}
[x_{i},p_{j}]=i\hbar\delta_{ij},\\{}
[p_{i},p_{j}]=0.\label{oc1}
\end{eqnarray}
On the basis of  (\ref{repx}), (\ref{repp}), the commutation relations for coordinates and momenta read
$[X_{i},P_{j}]=i\hbar\delta_{ij}+i\hbar\sum_k\theta_{ik}\eta_{jk}/{4}.$
So, parameters $\gamma_{ij}$ are considered in the following form
$\gamma_{ij}=\sum_k \theta_{ik}\eta_{jk}/4$ \cite{Bertolami}.

In noncommutative phase space of canonical type (\ref{form101})-(\ref{form10001}) the rotational symmetry is not preserved \cite{Chaichian,Balachandran1}. Algebra (\ref{form101})-(\ref{form10001}) is not rotationally invariant.  To preserve this symmetry  different types of noncommutative algebras were studied \cite{Moreno,Galikova,Amorim,GnatenkoPLA14}). Much attention has been devoted to studies of position-dependent noncommutativity  (see, for example, \cite{Lukierski,Lukierski2009,BorowiecEPL,Borowiec,Borowiec1,Kupriyanov2009,Kupriyanov}), noncommutative algebras with spin noncommutativity of coordinates (see, for example, \cite{Falomir09,Ferrari13,Deriglazov}).

In our previous paper \cite{GnatenkoIJMPA17} in order to construct rotationally invariant noncommutative algebra of canonical type we have studied the idea of involving additional coordinates and momenta. The parameters of noncommutativity were considered to be generalized to a tensors constructed with the help of the additional  coordinates and momenta.
In the present paper a system of interacting harmonic oscillators in uniform field is studied in the rotationally invariant noncommutative phase space. We investigate influence of coordinates noncommutativity and momentum noncommutativity on the spectrum of the system. On the basis of this result energy levels of a system of particles with harmonic oscillator interaction and a system of free particles in uniform field are analyzed in the rotationally invariant noncommutative phase space.

Nancommutative harmonic oscillator was studied in papers \cite{Hatzinikitas,Kijanka,Jing,Smailagic,Smailagic1,Muthukumar,Alvarez,Djemai,Dadic,Giri,Geloun,Abreu,Saha11,Nath,Shyiko}. A system  of two coupled harmonic oscillators was examined in two-dimensional noncommutative space \cite{Jellal,Jabbari}, in four-dimensional noncommutative phase space \cite{Bing,GnatenkoJPS17}. In \cite{BastosPhysA,Laba} system of free particles in four-dimensional noncommutative phase space of canonical type was studied.   In noncommutative space-time  classical problems of many particles among them $N$ interacting harmonic oscillators were examined in \cite{Daszkiewicz}.

  Studies of many-particle systems in the frame of noncommutative algebra of coordinates and momenta  give a possibility to find new effects in the properties of wide class of physical systems caused by space quantization.
Considered in the present paper system of interacting harmonic oscillators  has various applications.  Studies of a system of $N$ interacting harmonic oscillators are important  in nuclei physics \cite{Isgur,Glozman,Capstick}, in quantum chemistry and molecular spectroscopy \cite{Ikeda,Fillaux,Hong90,Michelot92}.  Recently networks of coupled harmonic oscillators have attracted much attention  because of their importance for quantum information processing \cite{Ponte2004,Ponte2007,Plenio}. The studies are also important for searching  signatures towards the Planck scale physics which are observable on macroscopic scales. The systems under consideration can be realized on the classical level as a system of oscillators coupled by springs.

Our paper is organized as follows. In Section 2  noncommutative algebra which is rotationally invariant and equivalent to noncommutative algebra of canonical type is presented. Section 3 is devoted to studies of the total hamiltonian in rotationally invariant noncommutative phase space. Effect of noncommutativity on the spectrum of a  system of N interacting harmonic oscillators in uniform field is examined in the Section 4. Also in this section a system of free particles in uniform filed and a system of particles with harmonic oscillator interaction are studied. Spectrum of a system of two interacting harmonic oscillators  and spectrum of three interacting harmonic oscillators are analyzed in Section 5 and Section 6, respectively. Section 7 is devoted to conclusions.

\section{Rotationally invariant noncommutative algebra of canonical type}

In our paper \cite{GnatenkoIJMPA17} we considered the tensors of noncommutativity to be defined as
\begin{eqnarray}
\theta_{ij}=\frac{c_{\theta} l^2_{P}}{\hbar}\sum_k\varepsilon_{ijk}\tilde{a}_{k}, \label{form130}\\
\eta_{ij}=\frac{c_{\eta}\hbar}{l^2_{P}}\sum_k\varepsilon_{ijk}\tilde{p}^b_{k}.\label{for130}
\end{eqnarray}
 where  $c_{\theta}$, $c_{\eta}$  are dimensionless constants, $l_P$ is the Planck length. We use notations $\tilde{a}_i$, $\tilde{b}_i$  $\tilde{p}^a_i$, $\tilde{p}^b_i$ for additional dimensionless coordinates and momenta conjugate to them which are governed by a spherically symmetric systems. For simplicity these systems are considered to be harmonic oscillators
 \begin{eqnarray}
 H^a_{osc}=\hbar\omega_{osc}\left(\frac{(\tilde{p}^{a})^{2}}{2}+\frac{\tilde{a}^{2}}{2}\right),\label{form104}\\
 H^b_{osc}=\hbar\omega_{osc}\left(\frac{(\tilde{p}^{b})^{2}}{2}+\frac{\tilde{b}^{2}}{2}\right).\label{for104}
 \end{eqnarray}
The values of parameters of noncommutativity are supposed to be of the order of the Planck scale. So, we put $\sqrt{{\hbar}}/\sqrt{{m_{osc}\omega_{osc}}}=l_{P}$. The frequency of the oscillators $\omega_{osc}$ is considered to be very large which leads to the statement that the oscillators put into the ground states remain in them \cite{GnatenkoIJMPA17}. So, in \cite{GnatenkoIJMPA17} we proposed the following noncommutative algebra
\begin{eqnarray}
[X_{i},X_{j}]=ic_{\theta} l^2_{P} \sum_k\varepsilon_{ijk}\tilde{a}_{k},\label{rotinv}\\{}
[X_{i},P_{j}]=i\hbar\left(\delta_{ij}+\frac{c_{\theta}c_{\eta}}{4}({\bf \tilde{a}}\cdot{\bf \tilde{p}^{b}})\delta_{ij}-\frac{c_{\theta}c_{\eta}}{4}{\tilde a}_j{\tilde p}^{b}_i\right),\\{}
[P_{i},P_{j}]=\frac{c_{\eta}\hbar^2}{l_P^2}\sum_k\varepsilon_{ijk}\tilde{p}^{b}_{k}.{}\label{rotinv1}
\end{eqnarray}
Commutation relations for  $\tilde{a}_i$, $\tilde{b}_i$  $\tilde{p}^a_i$, $\tilde{p}^b_i$  were considered to be as follows
 \begin{eqnarray}
[\tilde{a}_{i},\tilde{a}_{j}]=[\tilde{b}_{i},\tilde{b}_{j}]=[\tilde{a}_{i},\tilde{b}_{j}]=[\tilde{p}^{a}_{i},\tilde{p}^{a}_{j}]=\nonumber\\=[\tilde{p}^{b}_{i},\tilde{p}^{b}_{j}]=[\tilde{p}^{a}_{i},\tilde{p}^{b}_{j}]=0,\\{} [\tilde{a}_{i},\tilde{p}^{a}_{j}]=[\tilde{b}_{i},\tilde{p}^{b}_{j}]=i\delta_{ij},\\{}
[\tilde{a}_{i},\tilde{p}^{b}_{j}]=[\tilde{b}_{i},\tilde{p}^{a}_{j}]=0 {}\\ {} [\tilde{a}_{i},X_{j}]=[\tilde{a}_{i},P_{j}]=[\tilde{p}^{b}_{i},X_{j}]=[\tilde{p}^{b}_{i},P_{j}]=0.
 \end{eqnarray}
 So, like in the case of canonical version of noncommutativity with $\theta_{ij}$, $\eta_{ij}$, $\gamma_{ij}$ being constants, in the case of $\theta_{ij}$, $\eta_{ij}$ being defined as (\ref{form130}), (\ref{for130}) we can write
 \begin{eqnarray}
 [\theta_{ij}, X_k]=[\theta_{ij}, P_k]=[\eta_{ij}, X_k]=[\eta_{ij}, P_k]=[\gamma_{ij}, X_k]=[\gamma_{ij}, P_k]=0
 \end{eqnarray}
From this one can state that the proposed algebra
\begin{eqnarray}
[X_{i},X_{j}]=ic_{\theta} l^2_{P} \sum_k\varepsilon_{ijk}\tilde{a}_{k},\label{rotinv}\\{}
[X_{i},P_{j}]=i\hbar\left(\delta_{ij}+\frac{c_{\theta}c_{\eta}}{4}(\tilde{\bf {a}}\cdot\tilde{\bf {p}}^{b})\delta_{ij}-\frac{c_{\theta}c_{\eta}}{4}{\tilde a}_j{\tilde p}^{b}_i\right),\\{}
[P_{i},P_{j}]=\frac{c_{\eta}\hbar^2}{l_P^2}\sum_k\varepsilon_{ijk}\tilde{p}^{b}_{k},{}\label{rotinv1}
\end{eqnarray}
is equivalent to noncommutative algebra of canonical type at the same time it is rotationally invariant.

The noncommutative coordinates and noncommutative momenta which satisfy (\ref{rotinv})-(\ref{rotinv1}) can be represented as
\begin{eqnarray}
X_{i}=x_{i}+\frac{c_{\theta} l_P^2}{2\hbar}[\tilde{\bf {a}}\times{\bf p}]_i=x_{i}+\frac{1}{2}[{\bm \theta}\times{\bf p}]_i,\label{repx0}\\
P_{i}=p_{i}-\frac{c_{\eta}\hbar}{2l_P^2}[{\bf x}\times\tilde{\bf {p}}^b]_i=p_{i}-\frac{1}{2}[{\bf x}\times{\bm \eta}]_i,\label{repp0}
\end{eqnarray}
where coordinates and momenta $x_i$, $p_i$ satisfy (\ref{oc})-(\ref{oc1}) and for convenience the following vectors
\begin{eqnarray}
{\bm \theta}=(\theta_1,\theta_2,\theta_3), \ \ {\bm \eta}=(\eta_1,\eta_2,\eta_3),\\
\theta_i=\frac{1}{2}\sum_{jk}\varepsilon_{ijk}{\theta_{jk}},\\
\eta_i=\frac{1}{2}\sum_{jk}\varepsilon_{ijk}{\eta_{jk}},
\end{eqnarray}
 are introduced.

After rotation one has $X_{i}^{\prime}=U(\varphi)X_{i}U^{+}(\varphi)$, $P_{i}^{\prime}=U(\varphi)P_{i}U^{+}(\varphi)$ $a_{i}^{\prime}=U(\varphi)a_{i}U^{+}(\varphi)$,  $p^{b\prime}_{i}=U(\varphi)p^b_{i}U^{+}(\varphi)$. The rotation operator  $U(\varphi)=\exp(i\varphi({\bf n}\cdot{\bf L^t})/\hbar)$, contains the total angular momentum  which reads
${\bf L^t}=[{\bf x}\times{\bf p}]+\hbar[\tilde{\bf {a}}\times\tilde{\bf p}^{a}]+\hbar[\tilde{\bf { b}}\times\tilde{\bf {p}}^{b}]$ \cite{GnatenkoIJMPA17}.
The commutation relations  for coordinates and momenta remain the same
\begin{eqnarray}
[X^\prime_{i},X^\prime_{j}]=ic_{\theta} l^2_{P} \sum_k\varepsilon_{ijk}\tilde{a}^\prime_{k},\\{}
[X^\prime_{i},P^\prime_{j}]=i\hbar\left(\delta_{ij}+\frac{c_{\theta}c_{\eta}}{4}(\tilde{\bf {a}}^\prime\cdot\tilde{\bf {p}}^{b\prime})\delta_{ij}-\frac{c_{\theta}c_{\eta}}{4}{\tilde a}^\prime_j{\tilde p}^{b\prime}_i\right),\\{}
[P^{\prime}_{i},P^{\prime}_{j}]= \frac{c_{\eta}\hbar^2}{l_P^2}\sum_k\varepsilon_{ijk}\tilde{p}^{b\prime}_{k}.{}
\end{eqnarray}

\section{Hamiltonian of a system of interacting oscillators in noncommutative phase space with rotational symmetry}

Let us study a system of $N$ interacting harmonic oscillators of masses $m$ and frequencies $\omega$  in uniform field in noncommutative phase space with rotational symmetry (\ref{rotinv})-(\ref{rotinv1}).
The hamiltonian of the system reads
\begin{eqnarray}
 H_s=\sum_n\frac{( {\bf P}^{(n)})^{2}}{2m}+\sum_n\frac{m\omega^2( {\bf X}^{(n)})^{2}}{2}+\frac{k}{2}\mathop{\sum_{m,n}}\limits_{m\neq n}({\bf X}^{(n)}-{\bf X}^{(m)})^2+\nonumber\\+\kappa\sum_n X^{(n)}_1.\label{form777}
\end{eqnarray}
Here $\kappa$ and $k$ are constants.  The direction of the field for convenience is chosen to coincide with the $X_1$ axis direction. For $\kappa=0$, Hamiltonian (\ref{form777}) corresponds to nondissipative symmetric network of coupled harmonic oscillators \cite{Ponte2007}.

 Coordinates and momenta satisfy the following commutation relations
 \begin{eqnarray}
[X^{(n)}_{i},X^{(m)}_{j}]=i\hbar\delta_{mn}\theta^{(n)}_{ij},\label{ffor101}\\{}
[X^{(n)}_{i},P^{(m)}_{j}]=i\hbar\delta_{mn}\left(\delta_{ij}+\sum_k\frac{\theta^{(n)}_{ik}\eta^{(m)}_{jk}}{4}\right),\label{for1001}\\{}
[P^{(n)}_{i},P^{(m)}_{j}]=i\hbar\delta_{mn}\eta^{(n)}_{ij},\label{ffor10001}
\end{eqnarray}
with
\begin{eqnarray}
\theta^{(n)}_{ij}=\frac{c_{\theta}^{(n)}l_P^2}{\hbar}\sum_k\varepsilon_{ijk}\tilde{a}_{k}, \label{tn}\\
\eta^{(n)}_{ij}=\frac{c_{\eta}^{(n)}\hbar}{l_P^2}\sum_k\varepsilon_{ijk}\tilde{p}^b_{k},\label{etn}
 \end{eqnarray}
here indexes $m,n=(1...N)$ label the particles. Note that we consider the general case when different particles satisfy noncommutative algebra with different tensors of noncommutativity. The problem of description of composite system in rotationally invariant noncommutative phase space was discussed in our previous paper \cite{GnatenkoIJMPA18}. In the paper we proposed condition  on the parameters $c_{\theta}^{(n)}$, $c_{\eta}^{(n)}$ in tensors of noncommutativity on which the list of important results can be obtained (among them the noncommutative coordinates are independent on mass and noncommutative momenta are proportional to mass as it has to be,  coordinates and momenta of the center-of-mass commute with the coordinates and momenta of the relative motion \cite{GnatenkoIJMPA18}, the weak equivalence principle is recovered \cite{Gnatenko_arxiv}). The conditions read
\begin{eqnarray}
c^{(n)}_{\theta}m_n=\tilde{\gamma}=const, \ \ \frac{c^{(n)}_{\eta}}{m_n}=\tilde{\alpha}=const\label{conde}.
\end{eqnarray}
Constants $\tilde{\gamma}$, $\tilde{\alpha}$ are the same for particles with different masses. We would like also to note that the idea to relate parameters of  algebra for coordinates and momenta with mass is also important in deformed space with minimal length  \cite{Tk1,Tk2,Tk3}, two-dimensional noncommutative space of canonical type \cite{GnatenkoPLA13}, four-dimensional noncommutative phase space of canonical type \cite{GnatenkoPLA17,GnatenkoMPLA17}.

In the case of system of harmonic oscillators with  masses $m$ taking into account (\ref{tn}), (\ref{etn}), (\ref{conde}) one has
\begin{eqnarray}
\theta^{(n)}_{ij}=\theta_{ij}=\frac{c_{\theta} l_P^2}{\hbar}\sum_k\varepsilon_{ijk}\tilde{a}_{k}, \label{t}\\
\eta^{(n)}_{ij}=\eta_{ij}=\frac{c_{\eta}\hbar}{l_P^2}\sum_k\varepsilon_{ijk}\tilde{p}^b_{k},\label{e}
 \end{eqnarray}
with $c_{\theta}=\tilde{\gamma}/m$, $c_{\eta}=\tilde{\alpha} m$.

Using representation (\ref{repx0})-(\ref{repp0}) and  (\ref{t}), (\ref{e}) the hamiltonian of a system can be written in the following form
\begin{eqnarray}
 H_s=\sum_n\left(\frac{({\bf p}^{(n)})^{2}}{2m}+\frac{m\omega^2( {\bf x}^{(n)})^{2}}{2}+\kappa x^{(n)}_1\right)+\frac{k}{2}\mathop{\sum_{m,n}}\limits_{m\neq n}({\bf x}^{(n)}-{\bf x}^{(m)})^2+\nonumber\\+\sum_n\left(-\frac{({\bm \eta}\cdot{\bf L}^{(n)})}{2m}-\frac{m\omega^2({\bm \theta}\cdot{\bf L}^{(n)})}{2}+\frac{\kappa}{2}[{\bm \theta}\times {\bf p}^{(n)}]_1+\frac{m\omega^2}{8}[{\bm \theta}\times{\bf p}^{(n)}]^2+\right.\nonumber\\\left.+\frac{[{\bm \eta}\times{\bf x}^{(n)}]^2}{8m}\right)-
 \frac{k}{2}\mathop{\sum_{m,n}}\limits_{m\neq n}{\bm \theta}\cdot[({{\bf x}}^{(n)}-{{\bf x}}^{(m)})\times ({\bf p}^{(n)}-{\bf p}^{(m)})]+\nonumber\\+\mathop{\sum_{m,n}}\limits_{m\neq n} \frac{k}{8}[{\bm \theta}\times ({\bf p}^{(n)}-{\bf p}^{(m)})]^2,\label{orm777}
\end{eqnarray}
where ${\bf L}^{(n)}=[{\bf x}^{(n)}\times{\bf p}^{(n)}]$.
Because of involving of additional coordinates and additional momenta $\tilde{a}_i$, $\tilde{b}_i$  $\tilde{p}^a_i$, $\tilde{p}^b_i$ we have to consider the total hamiltonian which is the sum of $H_s$ and Hamiltonians of harmonic oscillators $H^a_{osc}$, $H^b_{osc}$
\begin{eqnarray}
H=H_s+H^a_{osc}+H^b_{osc}=H_0+\Delta H.\label{total1}
\end{eqnarray}
here
\begin{eqnarray}
H_0=\langle H_s\rangle_{ab}+H^a_{osc}+H^b_{osc},\label{2h0}\\
\Delta H= H-H_0=H_s-\langle H_s\rangle_{ab},
\end{eqnarray}
 $\langle...\rangle_{ab}$ denotes averaging over degrees of freedom of harmonic oscillators $H^a_{osc}$ $H^b_{osc}$ in the ground states
 \begin{eqnarray}
 \langle...\rangle_{ab}=\langle\psi^{a}_{0,0,0}\psi^{b}_{0,0,0}|...|\psi^{a}_{0,0,0}\psi^{b}_{0,0,0}\rangle
  \end{eqnarray}
$\psi^{a}_{0,0,0}$, $\psi^{b}_{0,0,0}$ are eigenstates of tree-dimensional harmonic oscillators $H^a_{osc}$, $H^b_{osc}$ in the ground states in the ordinary space (space with commutative coordinates and commutative momenta).

In our previous paper \cite{GnatenkoIJMPA18} we concluded that up to the second order in $\Delta H$ one can consider Hamiltonian $H_0$.

For a system of interacting harmonic oscillators using
 \begin{eqnarray}
 \langle\psi^{a}_{0,0,0}|\theta_i|\psi^{a}_{0,0,0}\rangle=\langle\psi^{b}_{0,0,0}|\eta_i|\psi^{b}_{0,0,0}\rangle=0, \\
\langle\theta_i\theta_j\rangle=\frac{c_{\theta}^2l_P^4}{\hbar^2}\langle\psi^{a}_{0,0,0}| \tilde{a}_i\tilde{a}_j|\psi^{a}_{0,0,0}\rangle=\frac{c_{\theta}^2l_P^4}{2\hbar^2}\delta_{ij}=\frac{\langle\theta^2\rangle\delta_{ij}}{3},\label{thetar2}\\
\langle\eta_i\eta_j\rangle= \frac{\hbar^2 c_{\eta}^2}{l_P^4}\langle\psi^{b}_{0,0,0}| \tilde{p}^{b}_i\tilde{p}^{b}_j|\psi^{b}_{0,0,0}\rangle=\frac{\hbar^2 c_{\eta}^2}{2 l_P^4}\delta_{ij}=\frac{\langle\eta^2\rangle\delta_{ij}}{3},\label{etar2}
\end{eqnarray}
and calculating
\begin{eqnarray}
\langle[{\bm \eta}\times{\bf x}^{(n)}]^2\rangle_{ab}=\frac{2}{3}\langle\eta^2\rangle({\bf x}^{(n)})^2, \ \ \langle[{\bm \theta}\times{\bf p}^{(n)}]^2\rangle_{ab}=\frac{2}{3}\langle\theta^2\rangle({\bf p}^{(n)})^2,\\
\langle[{\bm \theta}\times ({\bf p}^{(n)}-{\bf p}^{(m)})]^2\rangle_{ab}=\frac{2}{3}\langle\theta^2\rangle({\bf p}^{(n)}-{\bf p}^{(m)})^2
\end{eqnarray}
 the expression for $\Delta H$ can be written as
\begin{eqnarray}
\Delta H=\sum_n\left(-\frac{({\bm \eta}\cdot{\bf L}^{(n)})}{2m}-\frac{m\omega^2({\bm \theta}\cdot{\bf L}^{(n)})}{2}+\frac{\kappa}{2}[{\bm \theta}\times {\bf p}^{(n)}]_1+\frac{m\omega^2}{8}[{\bm \theta}\times{\bf p}^{(n)}]^2+\right.\nonumber\\\left.+\frac{[{\bm \eta}\times{\bf x}^{(n)}]^2}{8m}\right)-
 \frac{k}{2}\mathop{\sum_{m,n}}\limits_{m\neq n}{\bm \theta}\cdot[({{\bf x}}^{(n)}-{{\bf x}}^{(m)})\times ({\bf p}^{(n)}-{\bf p}^{(m)})]+\nonumber\\+\mathop{\sum_{m,n}}\limits_{m\neq n} \frac{k}{8}[{\bm \theta}\times ({\bf p}^{(n)}-{\bf p}^{(m)})]^2-\sum_n\left(\frac{\langle\eta^2\rangle({\bf x}^{(n)})^2}{12m}+\frac{\langle\theta^2\rangle m\omega^2({\bf p}^{(n)})^2}{12}\right)-\nonumber\\-\frac{k}{12}\mathop{\sum_{m,n}}\limits_{m\neq n}\langle{\theta}^2\rangle ({\bf p}^{(n)}-{\bf p}^{(m)})^2. \nonumber\\\label{delta}
 \end{eqnarray}
 So, on the basis of conclusion presented in \cite{GnatenkoIJMPA18} up to the second order in $\Delta H$ (or taking into account (\ref{delta}), up to the second order in the parameters of noncommutativity) for a system of interacting harmonic oscillators in uniform field we can consider the Hamiltonian as
 \begin{eqnarray}
H_0=\sum_n\left(\frac{({\bf p}^{(n)})^{2}}{2m}+\frac{m\omega^2( {\bf x}^{(n)})^{2}}{2}+\kappa x^{(n)}_1\right)+\frac{k}{2}\mathop{\sum_{m,n}}\limits_{m\neq n}({\bf x}^{(n)}-{\bf x}^{(m)})^2+\nonumber\\+ \sum_n\left(\frac{\langle\eta^2\rangle ({\bf x}^{(n)})^2}{12m}+\frac{\langle\theta^2\rangle m\omega^2({\bf p}^{(n)})^2}{12}\right)+\nonumber\\+\frac{k}{12}\mathop{\sum_{m,n}}\limits_{m\neq n}\langle{\theta}^2\rangle ({\bf p}^{(n)}-{\bf p}^{(m)})^2+H^a_{osc}+H^b_{osc}.\label{h00}
  \end{eqnarray}

\section{Influence of noncommutativity on the spectrum of a system of $N$ interacting oscillators}

Let us find energy levels of a system of $N$ interacting harmonic oscillators. It is convenient to introduce effective mass and effective frequency
 \begin{eqnarray}
 m_{eff}={m}\left({1+\frac{m^2\omega^2\langle\theta^2\rangle}{6}}\right)^{-1},\label{meff}\\
 \omega_{eff}=\left({\omega^2+\frac{\langle\eta^2\rangle}{6m^2}}\right)^{\frac{1}{2}}\left({1+\frac{m^2\omega^2\langle\theta^2\rangle}{6}}\right)^{\frac{1}{2}}\label{omegaeff}
   \end{eqnarray}
and rewrite (\ref{h00}) as
 \begin{eqnarray}
H_0=\sum_n\left(\frac{({\bf p}^{(n)})^{2}}{2m_{eff}}+\frac{m_{eff}\omega_{eff}^2( {\tilde{\bf x}}^{(n)})^{2}}{2}\right)-\frac{N\kappa^2}{2m_{eff}\omega^2_{eff}}+\nonumber\\+\frac{k}{2}\mathop{\sum_{m,n}}\limits_{m\neq n}({\tilde{\bf x}}^{(n)}-{\tilde{\bf x}}^{(m)})^2+\frac{k}{12}\mathop{\sum_{m,n}}\limits_{m\neq n}\langle{\theta}^2\rangle ({\bf p}^{(n)}-{\bf p}^{(m)})^2+H^a_{osc}+H^b_{osc}.\label{h5}
  \end{eqnarray}
where the vector
 \begin{eqnarray}
 \tilde{{\bf x}}^{(n)}=\left(x_1^{(n)}+\frac{\kappa}{m_{eff}\omega^2_{eff}},x^{(n)}_2,x^{(n)}_3\right),
  \end{eqnarray}
  is introduced.
 Coordinates and momenta ${\tilde{\bf x}}^{(n)}$, ${\bf p}^{(n)}$ satisfy the ordinary commutation relations
\begin{eqnarray}
[{\tilde x}^{(n)}_{i},{\tilde x}^{(m)}_{j}]=0,\label{noc}\\{}
[{\tilde x}^{(n)}_{i},p^{(m)}_{j}]=i\hbar\delta_{nm}\delta_{ij},\\{}
[p^{(n)}_{i},p^{(m)}_{j}]=0.\label{noc1}
\end{eqnarray}
 Note also that
 \begin{eqnarray}
[H_0,H^a_{osc}]=[H_0,H^b_{osc}]=0.
\end{eqnarray}
 Therefore, the spectrum of $H_0$ reads
\begin{eqnarray}
E_{\{n_1\},\{n_2\},\{n_3\}}=\sum^{N}_{a=1}\hbar\omega_a\left(n^{(a)}_1+n^{(a)}_2+n^{(a)}_3+\frac{3}{2}\right)-\frac{N\kappa^2}{2m_{eff}\omega^2_{eff}}+\nonumber\\+3\hbar\omega_{osc}.\label{en}
\end{eqnarray}
 where
  \begin{eqnarray}
 \omega_1=\omega_{eff},\label{omN}\\
 \omega_2=\omega_3=...=\omega_N=\nonumber\\=\left(\omega^2_{eff}+\frac{2kN}{m_{eff}}+\frac{kN\langle\theta^2\rangle m_{eff}\omega^2_{eff}}{3}+\frac{2k^2\langle\theta^2\rangle N^2}{3}\right)^{\frac{1}{2}}.\label{omN2}
 \end{eqnarray}
$n^{(a)}_i$ are quantum numbers ($n^{(a)}_i=0,1,2...$). In (\ref{en}) we take into account that the oscillators $H^a_{osc}$, $H^b_{osc}$ are in the ground states.
The first term in (\ref{en}) corresponds to the spectrum of the center-of-mass of the system. The terms with $a=2..N$ corresponds to the spectrum of the relative motion. This can be shown  introducing coordinates and momenta of the center-of-mass  ${\bf x}^c=\sum_n{\bf x}^{(n)}/N$, ${\bf p}^c=\sum_n {\bf p}^{(n)}$, and coordinates and momenta of relative motion $\Delta {\bf x}^{(n)}={\bf x}^{(n)}-{\bf x}^c$, $\Delta {\bf p}^{(n)}={\bf p}^{(n)}-{\bf p}^c/N$.  From (\ref{h5}) we can write
 \begin{eqnarray}
H_0=H^{c}+H_{rel}+H^a_{osc}+H^b_{osc},\label{tot32}\\
H^c=\frac{({\bf p}^c)^{2}}{2Nm_{eff}}+\frac{Nm_{eff}\omega_{eff}^2(\tilde{\bf x}^c)^2}{2}-\frac{N\kappa^2}{2m_{eff}\omega^2_{eff}},\\
H_{rel}=\sum_n\left(\frac{(\Delta{\bf p}^{(n)})^{2}}{2m_{eff}}+\frac{m_{eff}\omega_{eff}^2( \Delta{\bf x}^{(n)})^{2}}{2}\right)+\nonumber\\+\frac{k}{2}\mathop{\sum_{m,n}}\limits_{m\neq n}(\Delta{\bf x}^{(n)}-\Delta{\bf x}^{(m)})^2+\frac{k}{12}\mathop{\sum_{m,n}}\limits_{m\neq n}\langle{\theta}^2\rangle (\Delta{\bf p}^{(n)}-\Delta{\bf p}^{(m)})^2,\\{}
[H^{c},H_{rel}]=[H^{c},H^a_{osc}+H^b_{osc}]=[H_{rel},H^a_{osc}+H^b_{osc}]=0,{}
  \end{eqnarray}
where $\tilde{\bf x}^c=\left(x_1^{c}+{\kappa}/({m_{eff}\omega^2_{eff}}),x^{c}_2,x^{c}_3\right)$. So, from (\ref{en}) we have that the noncommutativity of coordinates and noncommutativity of momenta effects on the frequencies in the spectra of the center-of-mass and relative motion  of the system. The presents of uniform field shifts the spectrum on a constant.

In the limit $\langle\theta^2\rangle\rightarrow0$,  $\langle\eta^2\rangle\rightarrow0$ the expression for $E_{\{n_1\},\{n_2\},\{n_3\}}$ reduces to known spectrum for system of $N$ interacting harmonic oscillators in uniform field in the ordinary space
 \begin{eqnarray}
 E_{\{n_1\},\{n_2\},\{n_3\}}=\hbar\omega\left(n^{(1)}_1+n^{(1)}_2+n^{(1)}_3+\frac{3}{2}\right)+\nonumber\\+\sum^{N}_{a=2}\hbar\left(\omega^2+\frac{2Nk}{m}\right)^{\frac{1}{2}}\left(n^{(a)}_1+n^{(a)}_2+n^{(a)}_3+\frac{3}{2}\right)-\frac{N\kappa^2}{2m\omega^2}
\end{eqnarray}

From (\ref{en}) setting $\omega=0$, one can write the following expression for the spectrum of a system of $N$ particles of mass $m$ with harmonic oscillator interaction.
\begin{eqnarray}
E_{\{n_1\},\{n_2\},\{n_3\}}=\frac{\hbar\langle\eta^2\rangle}{6m^2}\left(n^{(1)}_1+n^{(1)}_2+n^{(1)}_3+\frac{3}{2}\right)+\nonumber\\+\hbar\left(\frac{2kN}{m}+\frac{\langle\eta^2\rangle}{6m^2}+\frac{2k^2\langle\theta^2\rangle N^2}{3}\right)^{\frac{1}{2}}\sum^{N}_{a=2}\left(n^{(a)}_1+n^{(a)}_2+n^{(a)}_3+\frac{3}{2}\right)-\nonumber\\-\frac{3N\kappa^2m}{\langle\eta^2\rangle}+3\hbar\omega_{osc}.\label{en1}
\end{eqnarray}
 The first term in (\ref{en1}) corresponds to the spectrum of the center-of-mass of the system.
In the contrast to the ordinary space (space with commutative coordinates and commutative momenta)  because of momentum noncommutativity the spectrum of the center-of-mass of the system of particles with harmonic oscillator interaction is discreet. The spectrum corresponds to the spectrum of harmonic oscillator with the frequency  ${\hbar\langle\eta^2\rangle}/{6m^2}$. The frequency in the spectrum of the relative motion is affected by the noncommutativity of coordinates and noncommutativity of momenta (see second term in (\ref{en1})).

It is worth  mentioning that from (\ref{en}) and (\ref{en1}) we have that the effect of coordinates noncommutativity on the spectrum of interacting harmonic oscillators (a system of particles with harmonic oscillator interaction) increases with increasing of the number of particles in the system.

For a system of $N$ free particles in uniform field in rotationally invariant noncommutative phase space  setting $k=0$ in (\ref{en1}) energy levels are as follows
\begin{eqnarray}
E_{\{n_1\},\{n_2\},\{n_3\}}=\sum^{N}_{a=1}\frac{\hbar\langle\eta^2\rangle}{6m^2}\left(n^{(a)}_1+n^{(a)}_2+n^{(a)}_3+\frac{3}{2}\right)-\frac{3N\kappa^2m}{\langle\eta^2\rangle}+3\hbar\omega_{osc}.\label{en2}
\end{eqnarray}
Note that the spectrum of a system of free particles is affected only by momentum noncommutativity. The spectrum corresponds to the spectrum of $N$ oscillators with frequencies determined by the parameter of momentum noncommutativity ${\hbar\langle\eta^2\rangle}/{6m^2}$.

We would like also to mention  that the presence of uniform field $\kappa$ shift the spectra  (\ref{en}), (\ref{en1}), (\ref{en2}) by constant.

\section{Two interacting oscillators in rotationally invariant noncommutative phase space}
Let us study particular case when a system consists of two oscillators with masses $m_1$, $m_2$ and frequencies $\omega_1$, $\omega_2$ and is described by the following Hamiltonian
\begin{eqnarray}
 H_s=\frac{( {\bf P}^{(1)})^{2}}{2m_1}+\frac{({\bf P}^{(2)})^{2}}{2m_2}+\frac{m_1\omega_1^2( {\bf X}^{(1)})^{2}}{2}+\frac{m_2\omega_2^2({\bf X}^{(2)})^{2}}{2}+k({\bf X}^{(1)}-{\bf X}^{(2)})^2.\label{2form777}
\end{eqnarray}
where ${\bf X}^{(n)}$, ${\bf P}^{(n)}$
satisfy  (\ref{ffor101})-(\ref{ffor10001}), ($n=1,2$).

System of two coupled harmonic oscillators has various applications in physics (see, for example, \cite{Jellal,Makarov} and references therein). The system is considered as a model in molecular physics \cite{Ikeda,Fillaux}, used for description of states of light in the framework of two-photon quantum optics \cite{Caves85,Schumaker85}.

 Taking into account (\ref{2h0}) and using representation (\ref{repx0})-(\ref{repp0}), we have
\begin{eqnarray}
H_0=\left(\frac{({\bf p}^{(1)})^{2}}{2m^{(1)}_{eff}}+\frac{({\bf p}^{(2)})^{2}}{2m^{(2)}_{eff}}+\frac{m^{(1)}_{eff}(\omega^{(1)}_{eff})^2( {{\bf x}}^{(1)})^{2}}{2}+\frac{m^{(2)}_{eff}(\omega^{(2)}_{eff})^2( {{\bf x}}^{(2)})^{2}}{2}\right)+\nonumber\\+k({{\bf x}}^{(1)}-{{\bf x}}^{(2)})^2+\frac{k}{6}\left(\langle({\theta^{(1)}})^2\rangle ({\bf p}^{(1)})^2+\langle(\theta^{(2)})^2\rangle ({\bf p}^{(2)})^2-\right.\nonumber\\\left.-2\langle{\theta^{(1)}\theta^{(2)}}\rangle({\bf p}^{(1)}\cdot{\bf p}^{(2)})\right)+H^a_{osc}+H^b_{osc}.
  \end{eqnarray}
with
   \begin{eqnarray}
 m^{(n)}_{eff}={m_n}\left({1+\frac{m_n^2\omega_n^2\langle(\theta^{(n)})^2\rangle}{6}}\right)^{-1},\label{2meff}\\
 \omega^{(n)}_{eff}=\left({\omega_n^2+\frac{\langle(\eta^{n})^2\rangle}{6m_n^2}}\right)^{\frac{1}{2}}\left({1+\frac{m_n^2\omega_n^2\langle(\theta^{(n)})^2\rangle}{6}}\right)^{\frac{1}{2}}\label{2omegaeff},\\
\langle\theta^{(n)}\theta^{(m)}\rangle=\frac{c^{(n)}_{\theta}c^{(m)}_{\theta}l_P^4}{\hbar^2}\langle\psi^{a}_{0,0,0}| \tilde{a}^2|\psi^{a}_{0,0,0}\rangle=\frac{3c^{(n)}_{\theta}c^{(m)}_{\theta}l_P^4}{2\hbar^2},\label{2thetar2}\\
\langle(\eta^{(n)})^{2}\rangle=\frac{\hbar^2 (c^{(n)}_{\eta})^2}{l_P^4}\langle\psi^{b}_{0,0,0}| (\tilde{p}^{b})^2|\psi^{b}_{0,0,0}\rangle=\frac{3\hbar^2 (c^{(n)}_{\eta})^2}{2 l_P^4},\label{2etar2}
\end{eqnarray}
Coordinates $x_i^{(n)}$ and momenta $p_i^{(n)}$ satisfy the ordinary commutation relations. So, the spectrum of $H_0$ is as follows
\begin{eqnarray}
E_{\{n_1\},\{n_2\},\{n_3\}}=\hbar{\omega}_+\left(n^{(1)}_1+n^{(1)}_2+n^{(1)}_3+\frac{3}{2}\right)+\nonumber\\+\hbar{\omega}_{-}\left(n^{(2)}_1+n^{(2)}_2+n^{(2)}_3+\frac{3}{2}\right)+3\hbar\omega_{osc}.\label{2en}
\end{eqnarray}
 where
  \begin{eqnarray}
 \omega^2_{\pm}=\frac{1}{{2}}\sum_n\left((\omega^{(n)}_{eff})^2+\frac{2k}{m^{(n)}_{eff}}+\frac{km^{(n)}_{eff}(\omega^{(n)}_{eff})^2\langle(\theta^{(n)})^{2}\rangle}{3}+\right.\nonumber\\\left.+\frac{2k^2}{3}\left(\langle(\theta^{(n)})^{2}\rangle+\langle\theta^{(1)}\theta^{(2)}\rangle\right)\right)\pm\frac{1}{2}\sqrt{D},
 \end{eqnarray}
 \begin{eqnarray}
 D=\left(\sum_n(\omega^{(n)}_{eff})^2+\sum_n\frac{2k}{m^{(n)}_{eff}}+\sum_n\frac{km^{(n)}_{eff}(\omega^{(n)}_{eff})^2\langle(\theta^{(n)})^{2}\rangle}{3}+\right.\nonumber\\ \left.+\sum_n\frac{2k^2}{3}\left(\langle(\theta^{(n)})^{2}\rangle+\langle\theta^{(1)}\theta^{(2)}\rangle\right)\right)^2-4\prod_n\left((\omega^{(n)}_{eff})^2+\frac{2k}{m^{(n)}_{eff}}+\right.\nonumber\\\left.+\frac{km^{(n)}_{eff}(\omega^{(n)}_{eff})^2\langle(\theta^{(n)})^{2}\rangle}{3}+\frac{2k^2}{3}\left(\langle(\theta^{(n)})^{2}\rangle+\langle\theta^{(1)}\theta^{(2)}\rangle\right)\right)+\nonumber\\
  +4\left(\frac{2k}{m^{(2)}_{eff}}+\frac{km^{(1)}_{eff}(\omega^{(1)}_{eff})^2\langle\theta^{(1)}\theta^{(2)}\rangle}{3}+\frac{2k^2}{3}\left(\langle(\theta^{(2)})^{2}\rangle+\langle\theta^{(1)}\theta^{(2)}\rangle\right)\right)\times\nonumber\\
  \left(\frac{2k}{m^{(1)}_{eff}}+\frac{km^{(2)}_{eff}(\omega^{(2)}_{eff})^2\langle\theta^{(1)}\theta^{(2)}\rangle}{3}+\frac{2k^2}{3}\left(\langle(\theta^{(1)})^{2}\rangle+\langle\theta^{(1)}\theta^{(2)}\rangle\right)\right).
 \end{eqnarray}
In the case of $m_1=m_2$ we have $m^{(n)}_{eff}=m_{eff}$, $\omega^{(n)}_{eff}=\omega_{eff}$ and the expressions reduce to
  \begin{eqnarray}
 \omega_-=\omega_{eff},\\
 \omega_+=\left(\omega^2_{eff}+\frac{4k}{m_{eff}}+\frac{2k\langle\theta^2\rangle m_{eff}\omega^2_{eff}}{3}+\frac{8k^2\langle\theta^2\rangle }{3}\right)^{\frac{1}{2}}.
 \end{eqnarray}
 which corresponds to (\ref{omN}), (\ref{omN2}) with $N=2$.

 \section{System of three interacting oscillators in rotationally invariant noncommutative phase space}
 Let us study a system of three interacting oscillators with masses $m_1$, $m_2=m_3=m$, and frequencies $\omega_1$, $\omega_2=\omega_3=\omega$. The Hamiltonian reads
 \begin{eqnarray}
 H_s=\frac{( {\bf P}^{(1)})^{2}}{2m_1}+\frac{( {\bf P}^{(2)})^{2}}{2m}+\frac{( {\bf P}^{(3)})^{2}}{2m}+\frac{m_1\omega_1^2( {\bf X}^{(1)})^{2}}{2}+\frac{m\omega^2({\bf X}^{(2)})^{2}}{2}+\nonumber\\+\frac{m\omega^2({\bf X}^{(3)})^{2}}{2}+k({\bf X}^{(1)}-{\bf X}^{(2)})^2+k({\bf X}^{(2)}-{\bf X}^{(3)})^2+k({\bf X}^{(3)}-{\bf X}^{(3)})^2.\label{3form777}
\end{eqnarray}
 In the case when $\omega_n=0$ the Hamiltonian (\ref{3form777})  is used as a model for description of confining forces between quarks \cite{Isgur,Glozman, Capstick}.
 Up to the second order in the parameters of noncommutativity one can consider
 \begin{eqnarray}
H_0=\sum_n\frac{({\bf p}^{(n)})^{2}}{2m^{(n)}_{eff}}+\sum_n\frac{m^{(n)}_{eff}(\omega^{(n)}_{eff})^2( {{\bf x}}^{(n)})^{2}}{2}+\nonumber\\+\frac{k}{2}\mathop{\sum_{m,n}}\limits_{m\neq n}({{\bf x}}^{(n)}-{{\bf x}}^{(m)})^2+\frac{k}{12}\mathop{\sum_{m,n}}\limits_{m\neq n}\left(\langle({\theta^{(n)}})^2\rangle ({\bf p}^{(n)})^2+\langle(\theta^{(m)})^2\rangle ({\bf p}^{(m)})^2-\right.\nonumber\\\left.-2\langle{\theta^{(n)}\theta^{(m)}}\rangle({\bf p}^{(n)}\cdot{\bf p}^{(m)})\right)+H^a_{osc}+H^b_{osc}.\label{3h5}
  \end{eqnarray}
where $m^{(n)}_{eff}$, $\omega^{(n)}_{eff}$, $\langle{\theta^{(n)}\theta^{(m)}}\rangle$ are defined as  (\ref{2meff})-(\ref{2thetar2}).

The spectrum of (\ref{3h5}) reads
\begin{eqnarray}
E_{\{n_1\},\{n_2\},\{n_3\}}=\sum^{3}_{a=1}\hbar\tilde{\omega}_a\left(n^{(a)}_1+n^{(a)}_2+n^{(a)}_3+\frac{3}{2}\right)+3\hbar\omega_{osc}.\label{3en}\\
\tilde{\omega}_1=\frac{1}{\sqrt{2}}\left(\omega^2_{eff}+(\omega^{(1)}_{eff})^2+\frac{2k}{m_{eff}}+\frac{4k}{m_{eff}^{(1)}}+A_1-\sqrt{D}\right)^{\frac{1}{2}},\label{om1}\\
\tilde{\omega}_2=\frac{1}{\sqrt{2}}\left(\omega^2_{eff}+(\omega^{(1)}_{eff})^2+\frac{2k}{m_{eff}}+\frac{4k}{m^{(1)}_{eff}}+A_1+\sqrt{D}\right)^{\frac{1}{2}},\label{om2}\\
\tilde{\omega}_3=\left(\omega_{eff}^2+\frac{6k}{m_{eff}}\right)^{\frac{1}{2}}\left(1+km_{eff}\langle {\theta}^2\rangle\right)^{\frac{1}{2}},\label{om3}
\end{eqnarray}
with
\begin{eqnarray}
D=\left(\omega^2_{eff}-(\omega^{(1)}_{eff})^2+\frac{4k}{m_{eff}}-\frac{4k}{m^{(1)}_{eff}}+A_2\right)^2+\left(\frac{2k}{m}+A_3\right)\left(2(\omega^{(1)}_{eff})^2-\right.\nonumber\\\left.-2\omega^2_{eff}-\frac{6k}{m}+\frac{8k}{m^{(1)}_{eff}}+8\left(\frac{2k}{m}+A_4\right)\left(\frac{2k}{m_1}+A_5\right)\left(\frac{2k}{m}+A_3\right)^{-1}+A_6\right),\\
A_1=\left(\frac{km_{eff}\omega^2_{eff}}{3}+\frac{2k^2}{3}\right)\langle {\theta}^2\rangle+\left(\frac{2km^{(1)}_{eff}(\omega^{(1)}_{eff})^2}{3}+\frac{8k^2}{3}\right)\langle (\theta^{(1)})^2\rangle+\nonumber\\+\frac{8k^2}{3}\langle\theta\theta^{(1)}\rangle,\\
A_2=\left(\frac{2km_{eff}\omega^2_{eff}}{3}+\frac{10k^2}{3}\right)\langle {\theta}^2\rangle-\left(\frac{2km^{(1)}_{eff}(\omega^{(1)}_{eff})^2}{3}+\frac{8k^2}{3}\right)\langle (\theta^{(1)})^2\rangle-\nonumber\\-\frac{2k^2}{3}\langle\theta\theta^{(1)}\rangle,\\
A_3=\left(\frac{8k^2}{3}+\frac{km_{eff}\omega_{eff}^{2}}{3}\right)\langle{\theta}^2\rangle-\frac{2k^2}{3}\langle\theta\theta^{(1)}\rangle,\\
A_4=\left(\frac{km^{(1)}_{eff}(\omega^{(1)}_{eff})^2}{3}+\frac{4k^2}{3}\right)\langle\theta\theta^{(1)}\rangle+\frac{2k^2}{3}\langle {\theta}^2\rangle,\\
A_5=\left(\frac{km_{eff}(\omega^2_{eff})}{3}+\frac{2k^2}{3}\right)\langle\theta\theta^{(1)}\rangle+\frac{4k^2}{3}\langle (\theta^{(1)})^2\rangle,\\
A_6=-\left({km_{eff}\omega^2_{eff}}+4k^2\right)\langle {\theta}^2\rangle+\left(\frac{4km^{(1)}_{eff}(\omega^{(1)}_{eff})^2}{3}+\frac{16k^2}{3}\right)\langle (\theta^{(1)})^2\rangle+\nonumber\\+\frac{2k^2}{3}\langle\theta\theta^{(1)}\rangle.
\end{eqnarray}
here for convenience we use notations $m_{eff}=m^{(2)}_{eff}=m^{(3)}_{eff}$, $\omega_{eff}=\omega^{(2)}_{eff}=\omega^{(3)}_{eff}$,  and $\theta=\theta^{(2)}=\theta^{(3)}$.

In the case when the masses and frequencies of the oscillators are equal, $m_1=m$, $\omega_1=\omega$, the result (\ref{3en}) reproduce (\ref{en}) with $N=3$. We have
\begin{eqnarray}
\tilde{\omega}_1=\omega_{eff},\\
\tilde{\omega}_2=\tilde{\omega}_3=\left(\omega^2_{eff}+\frac{6k}{m_{eff}}+k\langle\theta^2\rangle m_{eff}\omega^2_{eff}+6k^2\langle\theta^2\rangle\right)^{\frac{1}{2}}.
\end{eqnarray}

 For Hamiltonian  (\ref{3form777})  with $\omega_n=0$ which is considered for description of confining forces between quarks the spectrum is given by (\ref{3en}) with (\ref{om1}), (\ref{om2}), (\ref{om3}) and $m^{(1)}_{eff}=m_1$, $m_{eff}=m$, $\omega^{(1)}_{eff}=\sqrt{\langle(\eta^{1})^2\rangle}/\sqrt{6m_1^2}$, $\omega_{eff}=\sqrt{\langle(\eta)^2\rangle}/\sqrt{6m^2}$. Note, that because of noncommutativity of coordinates and noncommutativity of momenta the spectrum of the center-of-mass of the system is discrete and corresponds to the spectrum of harmonic oscillator with frequency $\tilde{\omega}_1$ (\ref{om1}).

 In a rotationally-invariant space with noncommutativity of coordinates (space which is characterized by (\ref{ffor101}), (\ref{for1001}) and $[P^{(n)}_{i},P^{(m)}_{j}]=0$), the spectrum of a system  described by Hamiltonian  (\ref{3form777})  with $\omega_n=0$ has the form (\ref{3en}) with frequencies
  \begin{eqnarray}
\tilde{\omega}_1=0,\label{som1}\\
\tilde{\omega}_2=\frac{1}{\sqrt{2}}\left(\frac{2k}{m}+\frac{4k}{m^{(1)}}+\frac{2k^2}{3}\langle {\theta}^2\rangle+\frac{8k^2}{3}\langle (\theta^{(1)})^2\rangle+\frac{8k^2}{3}\langle\theta\theta^{(1)}\rangle+\sqrt{D}\right)^{\frac{1}{2}},\label{som2}\\
\tilde{\omega}_3=\left(\frac{6k}{m}+6k^2\langle {\theta}^2\rangle\right)^{\frac{1}{2}},\label{som3}
\end{eqnarray}
where
\begin{eqnarray}
D=\left(\frac{4k}{m}-\frac{4k}{m^{(1)}}+\frac{10k^2}{3}\langle {\theta}^2\rangle-\frac{8k^2}{3}\langle (\theta^{(1)})^2\rangle-\frac{2k^2}{3}\langle\theta\theta^{(1)}\rangle\right)^2+\left(\frac{2k}{m}+\right.\nonumber\\\left.+\frac{8k^2}{3}\langle{\theta}^2\rangle-\frac{2k^2}{3}\langle\theta\theta^{(1)}\rangle\right)\left(-\frac{6k}{m}+\frac{8k}{m^{(1)}}+8\left(\frac{2k}{m}+\frac{4k^2}{3}\langle\theta\theta^{(1)}\rangle+\frac{2k^2}{3}\langle {\theta}^2\rangle\right)\times\right.\nonumber\\\times\left.\left(\frac{2k}{m_1}+\frac{2k^2}{3}\langle\theta\theta^{(1)}\rangle+\frac{4k^2}{3}\langle (\theta^{(1)})^2\rangle\right)\left(\frac{2k}{m}+\frac{8k^2}{3}\langle{\theta}^2\rangle-\frac{2k^2}{3}\langle\theta\theta^{(1)}\rangle\right)^{-1}-\right.\nonumber\\\left. -4k^2\langle {\theta}^2\rangle+\frac{16k^2}{3}\langle (\theta^{(1)})^2\rangle+\frac{2k^2}{3}\langle\theta\theta^{(1)}\rangle\right).\nonumber\\
\end{eqnarray}
 The frequencies are obtained putting $\omega_{eff}=\omega^{(1)}_{eff}=0$, $m^{(1)}_{eff}=m_1$, $m_{eff}=m$ in (\ref{om1}), (\ref{om2}), (\ref{om3}). Note, that  the spectrum of the center-of-mass of the system is not affected by noncommutativity of coordinates (\ref{som1}). The noncommutativity has influence on the frequencies in the spectrum of the relative motion  (\ref{som2}), (\ref{som3}).

\section{Conclusions}

We have considered noncommutative phase space of canonical type with rotational symmetry (\ref{rotinv})-(\ref{rotinv1}). The corresponding noncommutative algebra is constructed with the help of generalization of parameters of noncommutativity to tensors determined  by additional coordinates and additional momenta \cite{GnatenkoIJMPA17}.

In the frame of the rotationally invariant noncommutative algebra we have examined a system of $N$ harmonic oscillators with harmonic oscillator interaction in uniform field. The total hamiltonian has been constructed and analyzed (\ref{total1}). We have found energy levels of the system up to the second order in the parameters of noncommutativity.  We have obtained that noncommutativity  affects on the frequencies of the system (\ref{en}). Uniform field causes shift of the spectrum on a constant (\ref{en}). A system of two interacting oscillators and a system of three interacting oscillators have been studied in details and the corresponding spectra have been obtained (\ref{2en}), (\ref{3en}).

As particular cases, a system of particles with harmonic oscillator interaction  and  a system of free particles  have been studied in uniform field in rotationally-invariant noncommutative phase space. We have obtained that the spectrum of free particles in uniform field corresponds to the spectrum of a system of $N$ oscillators with frequencies ${\hbar\langle\eta^2\rangle}/{6m^2}$ and is not affected by the coordinates noncommutativity (\ref{en2}). For a system of  particles with harmonic oscillator interaction in uniform field we have found that the spectrum of the center-of-mass of the system is affected by noncommutativity of momenta and  corresponds to the spectrum of harmonic oscillator (see first term in (\ref{en1})). The spectrum of the relative motion of the system  corresponds to the spectrum of harmonic oscillators with frequencies determined by parameters of momentum noncommutativity and coordinate noncommutativity (see second term in (\ref{en1})).  We have concluded that effect of coordinates noncommutativity on the spectra of systems with harmonic oscillator interaction (system of interacting harmonic oscillators, system of particles with harmonic oscillator interaction) increases with increasing of the number of particles in the systems (\ref{en}), (\ref{en1}).

\section*{Acknowledgments}
The author thanks Prof. V. M. Tkachuk for his
advices and  support during research studies. This work was partly supported by the  by the State Found
for Fundamental Research under the project F-76.


\begin{thebibliography}{9}
\bibitem{Bertolami} O. Bertolami, R. Queiroz,  {\it Phys. Lett. A} {\bf 375}, 4116 (2011).
\bibitem{Chaichian} M. Chaichian, M. M. Sheikh-Jabbari, A. Tureanu, {\it Phys. Rev. Lett.} {\bf 86}, 2716 (2001).
\bibitem{Balachandran1} A. P. Balachandran, P. Padmanabhan,  {\it J. High Energy Phys.} {\bf 1012}, 001 (2010).


\bibitem{Moreno} E. F. Moreno, {\it Phys. Rev. D} {\bf 72}, 045001 (2005).
\bibitem{Galikova} V. G\'alikov\'a, P. Presnajder, {\it J. Phys: Conf. Ser.} {\bf 343}, 012096 (2012).

\bibitem{Amorim} R. Amorim, {\it Phys. Rev. Lett.} {\bf 101}, 081602 (2008).
\bibitem{GnatenkoPLA14} Kh.P. Gnatenko, V. M. Tkachuk, {\it Phys. Lett. A} {\bf 378}, 3509 (2014).

    \bibitem{Lukierski} M. Daszkiewicz, J. Lukierski, M. Woronowicz, {\it Phys. Rev. D} {\bf77}, 105007 (2008)
\bibitem{Lukierski2009} M. Daszkiewicz, J. Lukierski, M. Woronowicz, {\it J. Phys. A: Math. Theor.} {\bf42}, 355201 (2009).
\bibitem{BorowiecEPL} A. Borowiec, Kumar S. Gupta, S. Meljanac,  A. Pachol, {\it EPL } {\bf92}, 20006 (2010).
\bibitem{Borowiec} A. Borowiec, J. Lukierski, A. Pachol, {\it J. Phys. A: Math. Theor.} {\bf47} 405203 (2014).
\bibitem{Borowiec1} A. Borowiec, A. Pachol, {\it SIGMA } {\bf10}, 107 (2014).
\bibitem{Kupriyanov2009}  M. Gomes, V.G. Kupriyanov, {\it Phys. Rev. D} {\bf79}, 125011 (2009).
\bibitem{Kupriyanov} V. G. Kupriyanov, {\it J. Phys. A: Math. Theor.} {\bf 46}, 245303 (2013).

\bibitem{Falomir09} H. Falomir, J. Gamboa, J. López-Sarrión, F. Méndez, P.A.G. Pisani, {\it Phys. Lett. B}
{\bf680}  384 (2009).
\bibitem{Ferrari13} A.F. Ferrari, M. Gomes, V.G. Kupriyanov, C.A. Stechhahn, {\it Phys. Lett. B} {\bf718}, 1475 (2013).
\bibitem{Deriglazov} A. A. Deriglazov, A. M. Pupasov-Maksimov, {\it Eur. Phys. J. C }{\bf74}, 3101 (2014)


\bibitem{GnatenkoIJMPA17} Kh. P. Gnatenko, V. M. Tkachuk, {\it Int. J. Mod. Phys. A} {\bf32},  1750161  (2017).




\bibitem{Hatzinikitas} A. Hatzinikitas, I. Smyrnakis, {\it J. Math. Phys.} {\bf43},  113 (2002).
\bibitem{Kijanka} A. Kijanka, P. Kosinski, {\it Phys. Rev. D} {\bf70},  127702 (2004).
\bibitem{Jing} Jing Jian, Jian-Feng Chen, {\it Eur. Phys. J. C} {\bf60}, 669 (2009).
\bibitem{Smailagic} A. Smailagic, E. Spallucci, {\it Phys. Rev. D} {\bf65}, 107701 (2002).
\bibitem{Smailagic1} A. Smailagic, E. Spallucci, {\it J. Phys. A} {\bf35}, 363 (2002).
\bibitem{Muthukumar} B. Muthukumar, P. Mitra, {\it Phys. Rev. D} {\bf66}  027701 (2002).
\bibitem{Alvarez} P. D. Alvarez, J. Gomis, K. Kamimura, M. S. Plyushchay, {\it Phys. Lett. B} {\bf 659}  906 (2008).
\bibitem{Djemai} A. E. F. Djemai, H. Smail, {\it Commun. Theor. Phys.} {\bf41}, 837 (2004).
\bibitem{Dadic} I. Dadic, L. Jonke, S. Meljanac, {\it Acta  Phys. Slov.} {\bf 55}  149 (2005).
\bibitem{Giri}  P. R. Giri, P. Roy, {\it Eur. Phys. J. C} {\bf57}, 835 (2008).
\bibitem{Geloun} J. Ben Geloun,  S. Gangopadhyay, F. G. Scholtz, {\it EPL} {\bf86}, 51001 (2009).
\bibitem{Abreu}  E. M.C. Abreu, M. V. Marcial, A. C.R. Mendes,  W. Oliveira, {\it JHEP} 2013, 138 (2013).
\bibitem{Saha11}  A. Saha, S. Gangopadhyay, S. Saha, {\it Phys. Rev. D} {\bf83}, 025004 (2011)
\bibitem{Nath} D. Nath, P. Roy, {\it Ann. Phys}, {\bf377},  115  (2017).
\bibitem{Shyiko} Kh. P. Gnatenko, O. V. Shyiko, {\it Mod. Phys. Lett. A} {\bf33},  1850091  (2018).
\bibitem{Jellal} A. Jellal, El Hassan El Kinani, M. Schreiber, {\it Int. J. Mod. Phys. A} {\bf20}, 1515 (2005).
\bibitem{Jabbari} I.~Jabbari, A.~Jahan, Z~ Riazi, {\it Turk. J. Phys.} \textbf{33}, 149 (2009).
\bibitem{Bing} Bing-Sheng Lin, Si-Cong Jing, Tai-Hua Heng, {\it Mod. Phys. Lett. A} {\bf 23}, 445, (2008).
\bibitem{GnatenkoJPS17}  Kh. P. Gnatenko, V. M. Tkachuk, {\it J. Phys. Stud.} {\bf 21},  3001  (2017).
\bibitem{BastosPhysA} C. Bastos, A. E. Bernardini, J. F. G. Santos, {\it Physica A} {\bf438},  340 (2015).
\bibitem{Laba} Kh. P. Gnatenko, H. P. Laba, V. M. Tkachuk, {\it Mod. Phys. Lett. A} {\bf33},  1850131 (2018).
\bibitem{Daszkiewicz} M.~Daszkiewicz, C.J.~Walczyk, {\it Mod. Phys. Lett} A \textbf{26}, 819 (2011).
\bibitem{Isgur} N. Isgur, G. Karl, {\it Phys. Rev. D} {\bf 18}, 4187 (1978).
\bibitem{Glozman} L. Ya. Glozman, D.O. Riska, {\it Phys. Rept.} {\bf 268} 263, (1996)
\bibitem{Capstick} S. Capstick, W. Roberts  {\it Prog. Part. Nucl. Phys.} {\bf 45}, 241, (2000).
\bibitem{Ikeda} S. Ikeda, F. Fillaux, {\it Phys. Rev. B} {\bf 59} 4134 (1999).
\bibitem{Fillaux}  F. Fillaux, {\it Chem. Phys. Lett.} {\bf 408} 302306 (2005).
\bibitem{Hong90} Fan Hong-yi, {\it Phys. Rev. A}, {\bf42}, 4377 (1990).
\bibitem{Michelot92}  F. Michelot, {\it Phys. Rev. A} {\bf45}, 4271 (1992).
\bibitem{Ponte2004} M. A. de Ponte, M. C. de Oliveira, M. H. Y. Moussa, {\it Phys. Rev. A} {\bf 70} 022324, (2004).
\bibitem{Ponte2007}  M. A. de Ponte,S. S. Mizrahi, M. H. Y. Moussa, {\it Phys. Rev. A} {\bf76} 032101 (2007).
\bibitem{Plenio}   M.B. Plenio, J. Hartley, J. Eisert, {\it New J. Phys.} {\bf6}, 36 (2004)
\bibitem{Makarov}  D. N. Makarov, {\it Phys. Rev. E} {\bf97}, 042203 (2018).
\bibitem{Caves85} C.M. Caves and B.L. Schumaker, {\it Phys. Rev. A} {\bf 31}, 3068 (1985)
\bibitem{Schumaker85}  B.L. Schumaker and C.M. Caves, {\it Phys. Rev. A} {\bf 31}, 3093 (1985).

\bibitem{GnatenkoIJMPA18} Kh. P. Gnatenko, V. M. Tkachuk, {\it Int. J. Mod. Phys. A {\bf 33}},  1850037  (2018).
\bibitem{Gnatenko_arxiv} Kh. P. Gnatenko, arXiv:1808.00498.
\bibitem{Tk1}  V. M. Tkachuk, Phys. Rev. A {\bf86}, 062112 (2012).
\bibitem{Tk2}  C. Quesne, V.M. Tkachuk, {\it Phys. Rev. A} {\bf81}, 012106 (2010).
\bibitem{Tk3} V. M. Tkachuk, {\it Found. Phys.} {\bf46}, 1666 (2016).
\bibitem{GnatenkoPLA13} Kh. P. Gnatenko, {\it Phys. Lett. A} {\bf 377}, 3061 (2013).
\bibitem{GnatenkoPLA17} Kh. P. Gnatenko, V. M. Tkachuk, {\it Phys. Lett. A} {\bf381}, 2463 (2017).
\bibitem{GnatenkoMPLA17} Kh. P. Gnatenko, {\it Mod. Phys. Lett. A} {\bf32},  1750166 (2017).


\end{thebibliography}
\end{document}